\documentstyle[prb,twocolumn,aps]{revtex}
\tighten
\begin{document}
\draft
\title{Charging Ultrasmall Tunnel Junctions in Electromagnetic
Environment}
\author{X. H. Wang}
\address{Department of Theoretical Physics, Lund University,
S{\"o}lvegatan 14 A, S-223 62 Lund, Sweden}
\author{K. A. Chao}
\address{Department of Theoretical Physics, Lund University,
S{\"o}lvegatan 14 A, S-223 62 Lund, Sweden, and \\
Department of Physics, Norwegian University of Science and Technology,
N-7034 Trondheim, Norway}
\date{\today}
\maketitle
\begin{abstract}
We have investigated the quantum admittance of an ultrasmall tunnel
junction with arbitrary tunneling strength under an electromagnetic
environment. Using the functional integral approach a close analytical
expression of the quantum admittance is derived for a general
electromagnetic environment. We then consider a specific controllable
environment where a resistance is connected in series with the
tunneling junction, for which we derived the dc quantum conductance
from the zero frequency limit of the imaginary part of the quantum
admittance. For such electromagnetic environment the dc conductance
has been investigated in recent experiments, and our numerical results
agree quantitatively very well with the measurements. Our complete
numerical results for the entire range of junction conductance and
electromagnetic environmental conductance confirmed the few existing
theoretical conclusions.
\end{abstract}

\pacs{PACS numbers:  73.23.Hk, 85.30.Wx, 73.40.Rw}

\section{Introduction}
In semiclassical theory of Coulomb blockade\cite{alt91,c1}, electron
tunneling is controlled by the charging energy which is determined by
the junction capacitance. While the existence of this elementary
charging effect is clearly seen in multijunction configurations\cite{c2},
in a single junction, because of the stray capacitance, only partial
blockade has been observed.\cite{c3} The question is then, including
the electromagnetic environment, what is the effective junction
capacitance? The work of Devoret {\it et al\,}\cite{dev90} has shown that
external electrical circuit has strong impact on the effect of Coulomb
blockade in a single junction, and under standard experimental condition
where the external resistance is much smaller than the resistance
quantum $R_K$=$h/e^2$, the Coulomb gap is totally smeared out at low
voltages and at low temperature. A similar result was also obtained by
Girvin {\it et al}.\cite{gir90} In both works, as well as in most
existing studies on charging ultrasmall tunnel junctions, the tunnel
Hamiltonian was treated perturbatively\cite{alt91,gra92} using the
Fermi golden rule. This lowest order perturbative approach is valid for
weak tunneling strength.

With increasing tunneling strength, one must go beyond the Fermi golden
rule to include higher order tunneling processes. The conventional
perturbative calculation is very tedious, and has been used to study
the statistical properties of a single electron box
(SEB).\cite{but87,laf93,gra94} In a SEB an island is formed between a
tunnel junction and a gate capacitor. This is the simplest structure
exhibiting the charging effect. Even for such simplest system, it is
very difficult to derive the partition function beyond the second order
of the dimensionless tunnel conductance $R_K/R_T$, where $R_T$ is the
tunnel resistance.\cite{gra94} Therefore, an entirely different
approach is required to investigate the physics associated to strong
tunneling strength.

The functional integral method has been used to study an open, undriven
tunnel junction.\cite{amb82,ben83,sch90} As an extension of this work,
the relevant properties of a SEB have been investigated with the path
integral representation of the partition
function.\cite{sch90,pan91,fal95,wang96,wang97,weg} It has been
proved\cite{thesis} that a formal perturbation expansion of the
partition function results in exactly the same form as that derived
with the path integral method, provided that the channel number of the
tunnel junction is very large, which is the case for metallic tunnel
junctions. Furthermore, based on the functional integral formulism, a
Monte-Carlo simulation can be constructed for the entire regime from
weak to strong tunneling strength. The Monte-Carlo simulation has been
performed for the SEB and a very smooth interpolation between the weak
tunneling and the strong tunneling regime has been obtain.\cite{weg}
Therefore, functional integral is a reliable powerful tool to study
ultrasmall metallic tunnel junctions with arbitrary tunneling rate.

Since most experimental investigations on ultrasmall tunnel junctions
measure the current response to applied voltage, a relevant theoretical
study must take into account the influence of the electromagnetic
environment. This is the goal of the present work which will use the
function integral approach. In Sec.~II we introduce the model Hamiltonian
of a single tunnel junction under an electromagnetic environment in a
general form, and perform a theoretical analysis for arbitrary tunneling
rate. In Sec.~III we continue to derive the quantum admittance of such
system in a close analytical form. Then, in Sec.~IV we restrict
ourselves to the simple controllable electromagnetic environment which
allows us to compare our theoretical results with the recent
measurements.\cite{c4,c5} For this case we only need the dc conductance
which is the zero frequency limit of the imaginary part of the quantum
admittance. The numerical results will be presented in Sec.~V which
agree well with experiments.\cite{c4,c5} Finally, in Sec.~VI we give a
concluding remark.

\section{Theory for Model Hamiltonian}

The Hamiltonian $H_0$=$H_s$+$H_t$ of an open, undriven tunnel junction
consists of two parts. The first one
\begin{equation}
H_s=H_c+H_e \label{ho}
\end{equation}
represents the system in the absence of tunneling, where
\begin{equation}
H_c=Q^2/2C
\end{equation}
is the Coulomb charging energy when a charge $Q$ is added to the
junction with capacitance $C$. The electrodes are modelled as free
quasi-particle systems described by
\begin{equation}
H_e=\sum_{k,\sigma} \varepsilon_{k\sigma} c^+_{k\sigma} c_{k\sigma} +
\sum_{q,\sigma} \varepsilon_{q\sigma} c^+_{q\sigma} c_{q\sigma}, 
\end{equation}
where the {\it channel index} $\sigma$ includes both the transverse and
the spin quantum numbers. The longitudinal wave vectors in the left
electrod is labelled by $k$, and in the right electrod by $q$.
$c^+_{k\sigma}$ (or $c^+_{q\sigma}$) is the creation operator of an
electron with energy  $\varepsilon_{k\sigma}$ (or
$\varepsilon_{q\sigma}$) in the left (or right) electrode. The second
part of $H$ describes the tunneling processes of electrons
\begin{equation}\label{ka4}
H_t = \sum_{kq\sigma} [t_{kq\sigma}c^+_{k\sigma}
      c_{q\sigma} \exp (-i\varphi) +
      {\rm H.c.}] ,
\end{equation}
where $t_{kq\sigma}$ is the transition amplitude of an electron
tunneling from the state $q\sigma$ in the right electrode to the state
$k\sigma$ in the left one. During the tunneling the channel index
$\sigma$ is invariant. Here we have neglected the tunneling time, which
is approprite for metallic tunnel junctions. Otherwise the tunnel
Hamiltonian  must be modified.\cite{but85,but86,naz89} The operator
$\varphi$ is  conjugate to $Q$ via the commutation relation  
\[
[\varphi, Q]=-ie.
\]

To avoid possible ambiguity in our theoretical analysis, before
introducing the electromagnetice environment, let us first write down
the functional integral representation of the partition function of
$H_0$\cite{amb82,ben83}
\begin{equation}
Z_0=\int D\varphi e^{-S_0[\varphi(\tau)]} .
\end{equation}
In the action $S_0[\varphi]=S_c[\varphi]+S_t[\varphi]$,
\begin{equation}
S_c[\varphi]=\int_{0}^{\beta} d \tau \frac{{\dot{\varphi}}^2}
{4E_c}
\end{equation}
describes the kinetic part of the system, where $E_c=e^2/2C$ is the
elementary charging energy, and
\begin{equation}
S_t[\varphi]=-\int_{0}^{\beta} d\tau
\int_{0}^{\beta}  d\tau' \alpha_t(\tau-\tau') 
\cos [\varphi(\tau)-\varphi(\tau')]
\end{equation}
is resulted from the influence of tunneling processes. With the
commonly used approximation that the transition amplitude
$t_{kq\sigma}$ in (\ref{ka4}) is a constant $t_{kq\sigma}$=$t$, the
junction resistance $R_T$ can be calculated as
\[
R_T = (4\pi^2|t|^2\rho_l\rho_rN)^{-1} ,
\]
where $\rho_l$ (or $\rho_r$) is the density of state in the left (or
right) electrode, and $N$ is the number of channels. In this case the
damping kernel $\alpha_t(\tau)$ can be expressed as
\begin{equation}
\alpha_t(\tau) =
\frac{\alpha_t}{4\beta^2} \frac{1}{\sin^2 (\pi \tau/\beta)} ,
\end{equation}
where the coefficient $\alpha_t$=$R_K/R_T$ is the dimensionless
junction conductance. The Matsubara frequencies
$\omega_l=2\pi l/\beta$ which will be relevant to our future analysis
and calculation lies in the region much smaller than the bandwidth $D$
of the metal. For such frequencies the Fourier transform of
$\alpha_t(\tau)$ reduces to the simple form
\begin{equation}\label{ka7}
\alpha_t(\omega_l)=-\frac{\alpha_t |\omega_l|}{4\pi}.  
\end{equation}

The electromagnetic environment of the single tunnel junction
represented by $H_0$, caused by the on-chip leads and pads etc., will be
modelled by a transmission line, \cite{gir90} which can be described by 
a Caldeira-Leggett Hamiltonian\cite{dev90,gir90,cal83,leg84,wei93}
\begin{equation}
H_{\rm ex} = \sum_{n=1}^{\infty}
\left[ \frac{q_n^2}{2C_n} + \frac{\varphi_n^2}{2e^2 L_n} \right] 
\end{equation}
for an infinite series of LC circuits. In $H_{\rm ex}$ the first part
is the charging energy of all capacitors and the second part is the
magnetic energy of the corresponding inductors. The eigenfrequency of
the nth LC circuit is given by $\nu_n=1/\sqrt{L_n C_n}$, and the
spectral density  
\begin{equation}
J(\nu)=\pi \sum_{n=1}^{\infty} 
\frac{C_n \nu_n ^3}{2e^2} \delta(\nu-\nu_n)
\end{equation}
is determined by the external impedance. The operators $\varphi_n$ and
$q_n$  obey the commutation relation
\[
[\varphi_m, q_n]=-ie \delta_{m,n}.
\]
The electromagnetic environment $H_{\rm ex}$ is coupled to the tunnel 
junction $H_0$ bilinearly
\begin{equation}
H_{\rm int}=-\sum_{n=1}^{\infty} \frac{\varphi_n \varphi}{e^2 L_n}+
\sum_{n=1}^{\infty} \frac{\varphi^2}{2 e^2 L_n} .
\end{equation}
The second term at the right hand side of the above equation is a
physical counter term similar to the one introduced in the formulation
of the  dissipative quantum Brownian motion.\cite{wei93}

The partition function $Z$ of the total Hamiltonian
$H$=$H_0$+$H_{\rm ex}$+$H_{\rm int}$ can be similarly obtained in the
functional integral representation as
\begin{eqnarray}\label{ka1}
Z &=& \int D\varphi \prod_{n=1}^{\infty} \int D\varphi_n \\ 
& & \exp \left\{ -S_0[\varphi]-\int_{0}^{\beta} d\tau
\left[ \frac{C_n \dot{\varphi}_n^2} {2e^2}
+\frac{(\varphi-\varphi_n)^2}{2e^2 L_n} \right] \right\} . \nonumber
\end{eqnarray}
The path integrals over the bath modes are Gaussian, and thus can be 
evaluated exactly. From the condition of vanishing first variation of
the action with respect to $\varphi_n$, we obtain the classical 
equation of motion for the phase $\varphi_n$
\[
\ddot{\varphi}_n+\nu_n^2(\varphi-\varphi_n)=0.
\]
The solution to this linear differential equation with the boundary
conditions 
\[
\varphi_n(\beta)=\varphi_n(0)=\varphi_{n0}
\]
is
\begin{eqnarray}
& & \varphi_n^{\rm (cl)}(\tau) =
\frac{\sinh [\nu_n (\beta-\tau)]}{\sinh (\nu_n \beta)} \nonumber \\
& & \times \left[ \varphi_{n0} +
\frac{1}{\nu_n} \int_{0}^{\tau} d \tau' 
\sinh (\nu_n \tau') \varphi(\tau') \right] \nonumber \\
& & + \frac{\sinh (\nu_n \tau)}{\sinh (\nu_n \beta)} \nonumber \\
& & \times \left[ \varphi_{n0} -
\frac{1}{\nu_n} \int_{\tau}^{\beta} d \tau' 
\sinh [(\beta-\nu_n) \tau'] \varphi(\tau') \right] .
\end{eqnarray}  
The corresponding action is then calculated as
\begin{equation}\label{ka2}
S[\varphi_n^{\rm (cl)}]=S_{no}^{\rm (cl)}+S_{\rm ex}[\varphi] ,
\end{equation}
where $S_{no}^{\rm (cl)}$ depends only on the boundary value
$\varphi_{n0}$, and
\begin{equation}
S_{\rm ex}[\varphi] =
\frac{1}{2} \int_{0}^{\beta} d\tau \int_{0}^{\beta} d\tau'
\alpha_{\rm ex}(\tau-\tau') 
[\varphi(\tau)-\varphi(\tau')]^2 . \label{quas}
\end{equation}
The damping kernel of the electromagnetic environment
$\alpha_{\rm ex}(\tau)$ can be expressed in a Fourier series
\begin{equation}
\alpha_{\rm ex}(\tau) = \frac{1}{\beta}
\sum_{l=-\infty}^{\infty} 
\alpha_{\rm ex}(\omega_l) e^{i\omega_l \tau}
\end{equation}
with the Fourier coefficients
\begin{equation}
\alpha_{\rm ex}(\omega_l) = -\int_{0}^{\infty}
\frac{d \nu}{\pi}  \frac{J(\nu)}{\nu}
\frac{\omega_l^2}{\omega_l^2+\nu^2}
\end{equation}
related to the spectral density $J(\nu)$. $\alpha_{\rm ex}(\omega_l)$
is thus proportional to the Fourier tramsform of the admittance 
${\cal Y}_e(\omega)$ of the electromagnetic environment.\cite{leg84} 
In terms of the Matsubara frequences, this relation is simply
\begin{equation}\label{ka3}
\alpha_{\rm ex}(\omega_l) =
-\frac{R_K {\cal Y}_e(-i|\omega_l|)|\omega_l|}{4 \pi} .
\end{equation}

Both $S_{no}^{\rm (cl)}$ in (\ref{ka2}) and the contribution to
action from the fluctuations around classical paths are independent
of $\varphi$. Therefore, they will generate an irrelevant prefactor
in the partition function $Z$. We can neglect this factor and obtain
from (\ref{ka1}) the final result
\begin{equation}
Z=\int D\varphi e^{-S[\varphi]} ,
\end{equation}
with the total action 
\begin{equation}
S[\varphi]=S_0[\varphi]+S_{\rm ex}[\varphi] .
\end{equation}
We should point out that although $S_0[\varphi]$ is a periodic
function of $\varphi$, the total action $S[\varphi]$ is not. The
physical origin of the non-periodic feature of $S[\varphi]$ is the
continuous charge transfer between the two electrodes via the external
circuit, which suppresses the discrete nature of the change of charge
due to the tunneling processes. It is this modification of the system
geometry that can smear out dramatically the Coulomb charging effect,
as will be shown in the following sections.

\section{Quantum Admittance}

The quantum admittance $Y{(\omega)}$ of a single tunnel junction under
a general electromagnetic environment can be calculated conveniently
with the path integral approach by generalizing the Kubo formula for
this system~\cite{ben83,mah90,ho83,bro86}
\begin{equation}
Y(\omega) = \omega^{-1}
\left\{ \lim_{i\omega_l \rightarrow \omega+i\delta} 
\int_{0}^{\beta} d\tau e^{i\omega_l \tau}
\langle \hat{T}_\tau  I(\tau) I(0) \rangle \right\} , \label{cond}
\end{equation}
where $\hat{T}_\tau$ is the time-ordering operator in imaginary time.
The correlation function is calculated as
\begin{eqnarray}
& & \langle I(\tau)I(\tau') \rangle =Z^{-1} \int D\varphi
e^{-S[\varphi]} \{ 2e^2 \alpha_t(\tau-\tau') \nonumber \\
& & \times \cos [\varphi(\tau)-\varphi(\tau')]  
+I_T[\varphi, \tau] I_T[\varphi, \tau'] \} , \label{ii}
\end{eqnarray}
with the current functional
\begin{equation}
I_T[\varphi, \tau]=
-2e \int_{0}^{\beta} d\tau' \alpha_t(\tau-\tau')
\sin [\varphi(\tau)-\varphi(\tau')].
\end{equation}

At very low temperature one can use the instanton-like or the
renormalization group techniques to evaluate the partition
function\cite{fal95,wang96,wang97,kos76,sch94}, but not the
correlation function nor the quantum conductance, which are usually
calculated numerically.\cite{bro86} However, if the temperature is
not very low, with a semiclassical approximation we can evaluate
analytically all path integrals to obtain a systematic treatment of
the influence of fluctuations in the form of a series expansion of
$\beta E_c$. Another advantage of semiclassical method is that the
result is stable, because the fluctuation modes are of large
eigenvalues.\cite{wang96} Such relatively high temperature
semiclassical results of ultrasmall tunnel junctions turn out to be
very meaningful, in view of the recent measurements of Coulomb
blockade effects performed at not very low
temperature.\cite{c4,c5,pek94,est} In the following theoretical
analysis, we will use the semiclassical method.

Let us at first evaluate the partition function. The total action is
now approximated by a Gaussian form
\begin{eqnarray}
& & S_{\rm semi}[\varphi] =
\int_{0}^{\beta} d \tau \frac{{\dot{\varphi}}^2}{4E_c} \nonumber \\
& & +\frac{1}{2}\int_{0}^{\beta} d\tau
\int_{0}^{\beta}  d\tau' \alpha_w(\tau-\tau') 
[\varphi(\tau)-\varphi(\tau')]^2 , \label{gau}
\end{eqnarray}
where 
\begin{equation}
\alpha_w(\tau)=\alpha_t(\tau)+\alpha_{\rm ex}(\tau) .
\end{equation}
The above approximation means that the contribution of the 
tunneling resistance in the action is replaced by an equivalent 
Ohmic resistance.  Higher order variations of the action beyond the 
Gaussian approximataion can be calculated in the form of a power series 
of $\beta E_c$.  However, when calculating the current correlation function, 
which is normalized by $Z$, the power series correction factor in the 
denominator $Z$ is cancelled by similar factor in the numerator.  Hence, 
$S_{\rm semi}[\varphi]$ given by Eq.\ (\ref{gau}) turns out to be 
the approprite form for calculating the quantum admittance as a series of 
$\beta E_c$.  We notice that an arbitrary path $\varphi(\tau)$ may be 
expressed as a Fourier series
\begin{equation}
\varphi(\tau)=\sum_{l=-\infty}^{\infty} \varphi_l e^{i\omega_l \tau}
\label{Fphi}
\end{equation}
with the complex amplitudes 
\[
\varphi_l=\varphi_l'+i\varphi_l''
\]
under the condition $\varphi_{-l}=\varphi_l^*$. Hence, the
semiclassical action may be evaluated accordingly to give 
\begin{equation}
S_{\rm semi} [\varphi]=
\sum_{l=1}^{\infty} \lambda_l (\varphi_l'^2+\varphi_l''^2) ,
\end{equation}
where the eigenvalues $\lambda_l$ are
\begin{equation}\label{ka5}
\lambda_l = \frac{\omega_l^2}{2E_c} -
2[\alpha_t(\omega_l)+\alpha_{\rm ex}(\omega_l)] .
\end{equation}
Now the partition function can be directly calculated according to
the formula
\begin{equation}
Z_{\rm semi}=\prod_{l=1}^{\infty} \frac{\pi}{\beta \lambda_l}.
\end{equation} 

The current auto-correlation function (\ref{ii}) will be evaluated
in the same way. Let $\langle I(\tau)I(\tau')\rangle_1$ represent the
first term at the right hand side of (\ref{ii}). By performing a
series expansion in powers of $[\varphi(\tau)$-$\varphi(\tau')]$, we
can express $\langle I(\tau)I(\tau')\rangle_1$ as
\begin{eqnarray}
& & \langle I(\tau)I(\tau') \rangle_1 =
2e^2 \alpha_t(\tau-\tau') Z_{\rm semi}^{-1}
\int D\varphi e^{-S_{\rm semi}[\varphi]}  \nonumber \\
& & \times \left[ 1-\frac{[\varphi(\tau)-\varphi(\tau')]^2}{2}
+ \cdots \right].  
\end{eqnarray}
Making use of the Fourier series of paths $\varphi(\tau)$, we obtain
\begin{eqnarray}
& & \langle I(\tau)I(\tau') \rangle_1 =
2e^2 \alpha_t(\tau-\tau') \nonumber \\  
& & \times \left[1-2\sum_{n=1}^{\infty} 
\frac{1}{\beta \lambda_n} +
2\sum_{n=1}^{\infty}
\frac{\cos \omega_n(\tau-\tau')}{\beta \lambda_n} \right] \nonumber \\
& & +{\cal O} [(\beta E_c)^2].  
\end{eqnarray}
Substituting this expression of $\langle I(\tau)I(\tau') \rangle_1$
into (\ref{cond}), we obtain the contribution to the admittance from
the first term at the right hand side of (\ref{ii})
\begin{eqnarray}\label{g1}
& & Y_1(\omega) = 2e^2 \omega^{-1}
\left\{ \lim_{i\omega_l \rightarrow \omega+i\delta}
\right. \nonumber \\
& & \left. \left[ \left( 1-2\sum_{n=1}^{\infty}
\frac{1}{\beta \lambda_n}  \right)
\alpha_t(\omega_l) \right. \right. \nonumber \\ 
& & \left. \left.
+ \sum_{n=1}^{\infty} \frac{\alpha_t(\omega_l+\omega_n) 
+ \alpha_t(\omega_l-\omega_n)}{\beta \lambda_n}
\right] \right\} \nonumber \\
& & + {\cal O} [(\beta E_c)^2] .  
\end{eqnarray}
The second term on the right part of (\ref{ii}) can be evaluated in
the same way, although the algebra is more complicated. In terms of
the eigenvalues of the fluctuation modes and the Fourier tramsform of
the  damping kernel, its contribution to the admittance is derived as
\begin{eqnarray}\label{g2}
& & Y_2(\omega) = 4e^2 \omega^{-1} \left\{
\lim_{i\omega_l \rightarrow \omega+i\delta} \right. \nonumber \\
& & \left. \left[
\left( 1-4\sum_{n=1}^{\infty} \frac{1}{\beta \lambda_n} \right)
\frac{\alpha_t^2(\omega_l)}{\lambda_l} \right. \right. \nonumber \\ 
& & \left. \left. +\frac{2 \alpha_t(\omega_l)}{\lambda_l}
\sum_{n=1}^{\infty} \frac{\alpha_t(\omega_l+\omega_n) 
+\alpha_t(\omega_l-\omega_n)-2 \alpha_t(\omega_n)}{\beta \lambda_n} 
\right. \right. \nonumber \\
& & \left. \left. +\frac{8 \alpha_t^3(\omega_l)}{\lambda_l^2}
\sum_{n=1}^{\infty} \frac{1}{\beta \lambda_n}  
\right] \right\} \nonumber \\
& & + {\cal O} [(\beta E_c)^2] .
\end{eqnarray}

So far our theoretical analysis is for a general electromagnetic
environment. In the rest of this paper we will present a detail
quantitative study on dc conductance 
$G$$\equiv$$\lim_{\omega\rightarrow 0}\{ {\rm Im}\, Y(\omega)\}$ for
a specific case that the electromagnetic environment contains only an
Ohmic resistance. Such electromagnetic environment is relevant to
recent experiments\cite{c4,c5} with which our theoretical result can
be compared.

\section{Simple Electromagnetic Environment}

We will investegate the dc conductance of a single tunnel junction
connected in series with a resistance $R_{\rm ex}$. In real sample
this external resistance can be varied in a controllable way. In this
case we then have ${\cal Y}_e(-i|\omega_l|)=1/R_{\rm ex}$, and thus
(\ref{ka3}) becomes linear in frequency
\begin{equation}
\alpha_{\rm ex}(\omega_l)=-\frac{\alpha_{\rm ex} |\omega_l|}{4\pi} ,
\end{equation}
where $\alpha_{\rm ex}$=$R_K/R_{\rm ex}$ is the dimensionless
conductance contributed by the electromagnetic environment.
Substituting into (\ref{ka5}) this $\alpha_{\rm ex}(\omega_l)$ and
the $\alpha_t(\omega_l)$ given by (\ref{ka7}), we obtain the
eigenvelue $\lambda_l$ for the simple resistance environment. If $l$
is very large, $\lambda$ is dominated by the term $\omega_l^2/2E_c$,
and is insensitive to the tunneling terms represented by
$\alpha_{\rm ex}(\omega_l)$ and $\alpha_t(\omega_l)$. Hence, the
corresponding effect turns out to be an uninteresting prefactor and
contributes nothing to the conductance. For $\omega_l$ much less than
the bandwidth $D$, the eigenvalues reduce to
\begin{equation}
\lambda_l=\frac{\omega_l^2}{2E_c}+\frac{\alpha_w\omega_l}{2\pi} ,
\end{equation}
where $\alpha_w$=$\alpha_t$+$\alpha_{\rm ex}$. Inserting the
eigenvalues into (\ref{g1}), after a straightforward algebra, we  
obtain
\begin{eqnarray}
& & G_1 = \lim_{\omega\rightarrow 0}\{ {\rm Im}\,Y_1(\omega) \}
\nonumber \\
& & = 2e^2  \left\{
\left[ 1-\frac{2}{\alpha_w} [\Psi(\mu_w+1)-\Psi(1)] \right]
\frac{\alpha_t}{4\pi} \right. \nonumber \\ 
& & \left. -\frac{\mu_t \Psi'(\mu_w+1)}{2\pi} \right\}
+{\cal O} [(\beta E_c)^2] ,
\end{eqnarray}
where $\Psi(z)$ and $\Psi'(z)$ are, respectively, the digamma and
trigamma functions. The parameters $\mu_t$ and $\mu_w$ are defined
as
\begin{equation}
\mu_i = \alpha_i \beta E_c/2\pi^2  \,\,\,\,\, ; \, i=t,w .
\end{equation}

Now we calculate from (\ref{g2}) the second part of the dc
conductance $G_2$=$\lim_{\omega\rightarrow 0}\{ {\rm Im}Y_2(\omega)\}$.
When carrying out the limits at the right hand side of (\ref{g2}),
the orders in the series of $\beta E_c$ become mixed. This makes the
calculation of $G_2$ very complicated, but yet straightforwards.  The final 
result is 
\begin{eqnarray}
G_2 = &-& \frac{\alpha_t^2}{R_K \alpha_w}
\left\{ \alpha_w - 4 [\Psi(\mu_w+1)-\Psi(1)] \right\} \nonumber \\
&+& \frac{4 \mu_t \alpha_t}{R_K \alpha_w}
\Psi'(\mu_w + 1) \nonumber \\
&-& \frac{4 \alpha_t^3}{R_K \alpha_w^3} 
\left[ \Psi(\mu_w+1)-\Psi(1) \right] \nonumber \\ 
&+& {\rm O} [(\beta E_c)^2] .
\end{eqnarray}
Consequently, the dc conductance of a single tunneling junction under
the resistance environment is derived as
\begin{eqnarray}
G &=& \frac{1}{R_{\rm ex}+R_T} \left\{
1- \frac{\beta E_c}{\pi^2}
\frac{\alpha_{\rm ex}}{\alpha_w} \right. \nonumber \\ 
&\times& \left. \left[ \frac{\Psi(\mu_w+1)-\Psi(1)}{\mu_w}
+\Psi'(\mu_w+1) \right] \right. \nonumber \\
&+& \left. {\rm O} [(\beta E_c)^2] \right\} . 
\end{eqnarray}
In the high temperature limit, the tunneling system exhibits the 
conventional Ohmic behavior 
$G_{\rm ohm}=1/(R_T+R_{\rm ex})$ as expected.

\section{Results and discussion}

To demonstrate the influence of electromagnetic environment on Coulomb
charging effects, let us introduce the dimensionless conductance
correction
\begin{eqnarray}\label{dg}
& & \Delta g \equiv 1-G(R_{\rm ex}+R_T) \nonumber \\
& & = \frac{\beta E_c}{\pi^2} \frac{\alpha_{\rm ex}}{\alpha_w} 
\left[ \frac{\Psi(\mu_w+1)-\Psi(1)}{\mu_w}
+\Psi'(\mu_w+1)  \right] \nonumber \\
&+& {\rm O} [(\beta E_c)^2], 
\end{eqnarray}
which can be considered as a
measure of the effective Coulomb charging energy: smaller $\Delta g$
corresponds to weaker effective Coulomb charging energy. In recent
experiments\cite{c4,c5} Pekola and his co-workers have manufactured a
single tunnel junction with $\alpha_t$=$R_K/R_t$=0.5 and measured
1/$\Delta g$ as a function of normalized temperature $k_BT/E_c$ for
different values of $\alpha_{\rm ex}$=$R_K/R_{\rm ex}$. $R_{\rm ex}$
contains two contributions $R_{\rm ex}$=$R_{\rm ex,in}$+$R_{\rm ex,con}$,
where $R_{\rm ex,in}$ is the intrinsic value of the
circuit, which is of the order of the free space impedance 
$\sqrt{\mu_0/\varepsilon_0} \simeq $ 377 $\Omega$ and $R_{\rm ex,con}$ is 
experimentally controlable. At high
temperature 1/$\Delta g$ is found to be linear in $k_BT/E_c$.
Extrapolating this high temperature straight line to zero temperature,
it was found that the zero temperature offset $(1/\Delta g)_0$ is
positive and decreases with diminishing $\alpha_{\rm ex}$. Our theory
has reproduced these observed features as shown in Fig.~1 for
$\alpha_t$=$R_K/R_t$=0.5. For high temperature, 1/$\Delta g$ can be 
readily derived as 
\begin{equation}
\frac{1}{\Delta g}=\frac{3\alpha_w}{\alpha_{\rm ex}} \frac{k_B T}{E_c}
+\frac{(0.6+0.2\alpha_w)\alpha_w}{\alpha_{\rm ex}}.
\end{equation}
Thus, the offset is simply  
$(1/\Delta g)_0=(0.6+0.2\alpha_w)\alpha_w/\alpha_{\rm ex}$. 
For the case that the tunnel conductance $\alpha_t$ is much smaller than 
the electromagnetic environmental conductance $\alpha_{\rm ex}$, and 
in the vicinity of $\mu_w$$\ll$1, the offset
can be well approximated by the very simple form
$(1/\Delta g)_0$=$0.6+0.2R_K/R_{\rm ex}$. Hence, if we take 
$R_{\rm ex,in}$=377 $\Omega$,  we obtain
$(1/\Delta g)_0 \simeq$ 14 for $R_{\rm ex,con}$=0 and 
$(1/\Delta g)_0 \simeq$ 2 for
$R_{\rm ex,con}$=3 $k\Omega$, which agree exactly with the observed
values.\cite{c4,c5} An effective-capacitance model has been
proposed\cite{c4,c5} to explain these measured offset values. The
so-calculated $(1/\Delta g)_0$ is smaller than the experimental value
by a factor 5 for $R_{\rm ex,con}$=0, and fails to explain the case of
$R_{\rm ex,con}$=3 $k\Omega$. 

For a fixed value of $\beta E_c$, $\Delta g$ is a function of the
junction conductance $\alpha_t$ and the electromagnetic environmental
conductance $\alpha_{\rm ex}$. Using a standard formalism, it has been 
proven \cite{est} that for
high temperature, in the plane of $\alpha_t$ and $\alpha_{\rm ex}$,
the ratio $\Delta g/\beta E_c$ has an absolute maximum value $1/3$.
However, the functional dependence of $\Delta g/\beta E_c$ on
$\alpha_t$ and $\alpha_{\rm ex}$ has never been obtained. Our theory
allows us to calculate such important results which are shown in Fig.~2
for $\beta E_c$=0.2 and in Fig.~3 for $\beta E_c$=0.02. In each figure
the ratio $\pi^2\Delta g/\beta E_c$ is plotted as a function of
$\alpha_{\rm ex}$ for $\alpha_t$=0.1 (solid curve), 1.0 (dot-dashed
curve), 10.0 (dotted curve), and 100.0 (dashed curve). It is clear that
in the $\alpha_t$-$\alpha_{\rm ex}$ plane, the absolute maximum value of
$\Delta g/\beta E_c$ lies in the region where
$R_T$$\gg$$R_{\rm ex}$$\geq R_K$. A careful numerical search reveals
that the value of this absolute maximum is just $1/3$, corresponding
to the charging effects due to the geometric capacitance of the
ultrasmall tunnel junction.

We notices that in both Fig.~2 and Fig.~3, for a fixed value of
$\alpha_{\rm ex}$, the value of $\pi^2\Delta g/\beta E_c$ decreases
monotonically with $\alpha_t$. This is understandable because an
increase of the junction conductance (or the tunnel strength) leads to
a reduction of the effective Coulomb charging energy. On the other
hand, for a fixed value of the junction conductance $\alpha_t$,
$\pi^2\Delta g/\beta E_c$ as a function of the external conductance
$\alpha_{\rm ex}$ exhibits a peak. Such interesting feature has been
observed very recently, but has not appeared in literature
yet.\cite{c5} The physics of this peak structure can be explained as
follow. Once an electron tunnels through the junction from the left side
to the right side, it can return to the left side via the external
circuit. Therefore, the larger is the external conductance
$\alpha_{\rm ex}$, the faster can the tunnel junction {\em relax} to its
initial charge configuration. Consequently, the effective Coulomb
charging energy is reduced and so is $\Delta g$. This conclusion then
agrees with the one reached in Ref.\onlinecite{dev90} and
Ref.\onlinecite{gir90}. It is worthwhile to mention that in the limit of
very high tunneling rate or very large $\alpha_{\rm ex}$ such that
$\mu_w$$\gg$1, (\ref{dg}) is simplified to
\begin{equation}
\Delta g = \frac{2\alpha_{\rm ex}}{\alpha_w^2} \left[
\ln \{ \frac{\alpha_w\beta E_c}{2\pi^2} \} + 1 - \Psi(1)  \right] ,
\end{equation}
which is not analytical in $\alpha_t$. Therefore, the method fo
calculating $G$ in Ref.\onlinecite{but86} based on the Fermi golden
rule is no longer valid here.  

The picture is entirely different at the other end of very small
external conductance $\alpha_{\rm ex}$, or very large external
resistance $R_{\rm ex}/R_K$. In this case the external returning path
from the right side of the junction to the left side is heavily
blocked. Hence, once an electron tunnels through the junction, its
probability to tunnel back is much larger than that to travel through
the external circuit. Eventually, an electron spends most of the time
tunneling back and forth, and so higher order tunneling terms must be
taken into account, even though each tunnel process is still
incoherently sequential. Such tunneling processes cannot be treated
with the Fermi golden rule neither. In fact, the system is in a
dynamical state with an equivalent reduced effective Coulomb charging
energy. Each curve in Fig.~2 and Fig.~3 then drops down to zero as
$\alpha_{\rm ex}$ approaches zero, at which the external circuit is
completely blocked. This point of view is similar to that appeares in
the investigation of the statistical behavior of a single electron
box.\cite{gra94,pan91,fal95,wang96,wang97,weg,thesis,sch94}
Nevertheless, in the present problem the renormalization of the
capacitance is dynamical rather than static. To our knowledge, such
dynamical renormalization of the charging energy in the regime of very
small external conductance $\alpha_{\rm ex}$ is discovered for the
first time.

\section{Remarks}

We have used a non-perturbative approach to derive the quantum
admittance of a single tunnel junction in the presence of an
electromagnetic environment. Besides the case with an Ohmic environment
which has been investigated in the present work, our theoretical
formulation is very general and can be conveniently used to study the
effects of non-Ohmic environments. If the electromagnetic environment
contains a frequency-dependent part described by a transmission
line\cite{dev90,gir90,leg84,cha86}, the quantum admittance of the
system as a function of frequency, which is of great interesting, can
be analyzed with our theory. Such theoretical results can be compared
with experiments in a very direct manner.\cite{c5,but96} With a slight
modification of the action, we can also investigate the behavior of an
array of tunnel junctions. If the junction array has a pure Ohmic
environment, the dc conductance of the system can be calculated by
evaluating the mean value of the current through the external
resistance.\cite{wangb} Finally, we should mention that very recently
a dissipation-driven superconductor-insulator transition in a Josephson
junction array has been observed by changing the external impedance
continuously.\cite{rim97} Using the theoretical approach developed in
the present work, this problem can be investigated theoretically at a
quantitative level.

\section*{ACKNOWLEDGMENTS}
This work was supported by the Swedish Natural Science Research Council. 
The authors would like to thank J.\ P.\ Pekola for providing preliminary
experimental results and stimulating discussions, and to thank Yu.\ M.\ 
Galperin and P.\ Johansson for fruitful discussions about the theoretical 
aspect of this work.


\begin{figure}
\caption{$1/\Delta g$ as a function of the normalized temperature
$k_BT/E_c$ for the dimensionless junction conductance $\alpha_t$=0.5,
with various values of the dimensionless external conductance
$\alpha_{\rm ex}$=10 (solid curve), 50 (dotted curve), 100 (dashed curve), 
and 150 (dot-dashed curve). The cases $\alpha_{\rm ex}$=10 and 100
correspond closely to the two measured samples in Ref.~20 and Ref.~21,
respectively.}
\label{fig1}
\end{figure}

\begin{figure}
\caption{$\Delta g$ in units of $\beta E_c/\pi^2$ as a function of
the dimensionless external conductance $\alpha_{\rm ex}$ for
$\beta E_c$=0.2, with various values of the dimensionless junction
conductance $\alpha_t$=0.1 (solid curve), 1.0 (dot-dashed curve), 10.0
(dotted curve) and 100.0 (dashed curve).}
\label{fig2}
\end{figure}

\begin{figure}
\caption{The same as Fig.~2 but for $\beta E_c$=0.02.}
\label{fig3}
\end{figure}

\end{document}